\documentclass{PoS}
\usepackage[utf8]{inputenc}
\usepackage{amsmath}
\usepackage{amsfonts}
\usepackage{amssymb}
\usepackage{graphicx}
\usepackage{capt-of}

\title{Constraining the Doublet Left-Right Model}

\ShortTitle{Constraining the Doublet Left-Right Model}

\author{\speaker{Luiz Vale Silva}%
		\thanks{In collaboration with S. Descotes-Genon (LPT) and V. Bernard (IPNO). This project is partially supported by P2IO (Physique des 2 Infinis et des Origines) and CNRS.}\\
		Laboratoire de Physique Th\'{e}orique (LPT), CNRS/Univ. Paris-Sud 11 (UMR 8627) and
		
		Institut de Physique Nucl\'{e}aire d'Orsay (IPNO), CNRS/IN2P3/Univ. Paris-Sud 11 (UMR 8608)\\
		E-mail: \email{Luiz.Vale@th.u-psud.fr}}

\abstract{Left-Right Models (LRM) attempt at giving an understanding of the violation of parity (or charge-conjugation) by the weak interactions in the SM through a similar description of left- and right-handed currents at high energies. The spontaneous symmetry breaking of the LRM gauge group is triggered by an enlarged Higgs sector, usually consisting of two triplet fields (left-right symmetry breaking) and a bidoublet (electroweak symmetry breaking). I reconsider an alternative LRM with doublet instead of triplet fields. After explaining some features of this model, I discuss constraints on its parameters using electroweak precision observables (combined using the \textsf{CKMfitter} frequentist statistical framework) and neutral-meson mixing observables.}

\FullConference{Flavorful Ways to New Physics\\
				28-31 October 2014\\
				Freudenstadt - Lauterbad, Germany}
\begin{document}

% Starter plus problematic
One of the most puzzling features of the Standard Model (SM) consists in the different treatment of left and right chiralities 
of fermions, as shown by the violation of parity by weak interactions. In order to restore this symmetry at high energies, Left-Right Models (LRM) have been introduced in the 70's \cite{History} and they assume a symmetry between left- and right-handed fermions broken spontaneously, implying that left- and right-handed currents behave differently at low energies.

% Historical context, problems in the model and our model
Historically, LRM have been considered with doublets in order to break the left-right symmetry spontaneously \cite{History}. Later the focus was on triplet models, due to their ability to generate both Dirac and Majorana masses for neutrinos. The triplet models have the advantage of potentially introducing a see-saw mechanism, though it is difficult to reconcile the very light masses of neutrinos with a TeV scale LRM without fine tuning \cite{Deshpande}. Moreover, combined constraints coming from meson oscillations, among other observables, tend to push the mass scale of the new scalar particles to a few TeV \cite{Buras} or beyond \cite{Mohapatra}. The new vector particles must not be far away from this scale, otherwise the couplings present in the Higgs potential would become non-perturbative. Much effort has been done to avoid these constraints, but stringent lower bounds persist \cite{Basecq}.

Our aim is to reconsider the breaking of the left-right gauge group, via doublet rather than triplet fields, and see the constraints set on the scale and pattern of symmetry breaking. We also want to determine whether experimental data can be accommodated only through this spontaneous breakdown or if it requires also an explicit breaking of parity through different couplings in the left and right sectors.

\section{Doublet LRM}

% Setting the model, gauge group % Doublet, scale of the NP, new gauge bosons
The gauge group of LRM is $ SU(3)_{c} \times SU(2)_{L} \times SU(2)_{R} \times U(1)_{B-L} $, where $ B $ is the baryon number and $ L $ is the lepton number. We call $ g_{B-L} $ the gauge coupling of $ U(1)_{B-L} $, and $ g_{L} $ and $ g_{R} $ the ones of $ SU(2)_{L} $ and $ SU(2)_{R} $, respectively -- the case $ g_{L} \neq g_{R} $ explicitly violates parity. The LR symmetry is spontaneously broken into the EW symmetry, $ SU(3)_{c} \times SU(2)_{L} \times U(1)_{Y} $, where $ Y $ is the hypercharge, given by $ Y = T^{3}_{R} + \frac{B-L}{2} $. This breaking can be triggered by a scalar in any representation whose Vacuum Expectation Value (VEV) does
not preserve the LR symmetry, but preserves the EW symmetry. One considers here a doublet
representation $ \chi_{R} = \begin{pmatrix}
\chi^{\pm}_{R} , \chi^{0}_{R}
\end{pmatrix} $, with quantum numbers under the gauge group $ (0,0,1/2,1) $, whose VEV is $ \langle \chi_{R} \rangle = \begin{pmatrix}
0 , \frac{1}{\sqrt{2}} \kappa_{R}
\end{pmatrix} $. Since LRM is assumed to be valid at energies much higher than the scale $ \kappa $ of the EW Symmetry Breaking (EWSB), it follows that $ \kappa_{R} \gg \kappa $. The Higgs mechanism then leads to new heavy gauge bosons, called $ {W'}^{\pm} $ and $ {Z'}^{0} $, whose masses are of order $ \kappa_{R} $, coupled predominantly to right-handed fermions.

% Bi-doublet and freedom in the second doublet
To discuss the physics occurring at the EWSB, one introduces a bidoublet $ \phi = \begin{pmatrix}
\varphi^{0}_{1} & \varphi^{+}_{2} \\
\varphi^{-}_{1} & \varphi^{0}_{2} \\
\end{pmatrix} $, $ (0,1/2,1/2,0) $, whose VEV, $ \langle \phi \rangle = \frac{\operatorname{diag} \left( \kappa_{1}, \kappa_{2} \right)}{\sqrt{2}} $, is not invariant under the SM gauge group and breaks it spontaneously into $ SU(3)_{c} \times U(1)_{\operatorname{EM}} $. Even though a second doublet $ \chi_{L} $, with quantum numbers $ (0,1/2,0,1) $, is not necessary from the point-of-view of EWSB, it is introduced in order to preserve the structural symmetry between left and right sectors before symmetry breaking. Its VEV is $ \langle \chi_{L} \rangle = \begin{pmatrix}
0 , \frac{1}{\sqrt{2}} \kappa_{L}
\end{pmatrix} $, with $ \kappa_{L} $ of the order of the EWSB scale at most, and thus it also triggers the EWSB. Since $ \chi_{L} $ is a doublet, its VEV does not contribute to the $ \rho $ parameter at tree-level, and must be constrained by other observables, in particular EW Precision Observables (EWPO). The scale of EWSB is set by $ \kappa^{2} \equiv \kappa^{2}_{1} + \kappa^{2}_{2} + \kappa^{2}_{L} $ and the SM Higgs is given by a combination of the real degrees of freedom of $ \phi $ and $ \chi_{L} $. The gauge bosons $ {W}^{\pm} $ and $ {Z}^{0} $ acquire masses of order $ \kappa $ by the Higgs mechanism, and they couple predominantly to left-handed fermions. For simplicity, we take $ \kappa_{1,2} $ and $ \kappa_{L,R} $ to be real and positive. We also consider that there is no complex phase in the Higgs potential, so that no new CP-violation terms are generated by the extended Higgs sector.

% Yukawa
In the LRM, right-handed (left-handed) fermions come into doublets (singlets) of $ SU(2)_{R} $ and singlets (doublets) of $ SU(2)_{L} $, denoted $ Q_{R} $ ($ Q_{L} $). The mechanism responsible for giving them a mass is the Yukawa coupling $ \overline{Q}_{L} \left( Y \phi + \tilde{Y} \sigma_{2} \phi^{*} \sigma_{2} \right) Q_{R} + h.c. $, where generation indices are not shown. As in the SM, one introduces the mixing matrices $ V^{L,R} $, where $ V^{L} $ is the equivalent of the SM-CKM matrix and $ V^{R} $ is a new mixing matrix for right-handed quarks. Discrete symmetries can be imposed to relate L and R sectors, implying relations between $ V^{L} $ and $ V^{R} $ \cite{Maiezza}. The more general case where $ V^{L,R} $ are independent corresponds to an explicit violation of parity.

% h0,... and masses
The spectrum of physical scalars of the Doublet LRM is composed of one light neutral Higgs $ h^{0} $, five heavy neutral Higgses $ H^{0}_{1,2,3} $ (CP-even) and $ A^{0}_{1,2} $ (CP-odd), and four heavy charged Higgses $ H^{\pm}_{1,2} $. All of the heavy Higgses have masses of the order of $ \kappa_{R} $. The neutral Higgses $ H^{0}_{1,2} $ and $ A^{0}_{1,2} $ couple to quarks with a strength proportional to the Yukawa couplings and VEV's and induce Flavor Changing Neutral Currents (FCNC), whereas charged Higgses induce left- and right-handed Flavor Changing Charged Currents (FCCC).

Some differences/advantages of the doublet model compared to the triplet case are the following: (1) A VEV, whose size (of the order of the EWSB or less) is less constrained than in the triplet model, modifies the structure of FCNC couplings between Higgses and quarks; (2) This impacts the analysis of neutral-meson mixing (mainly corrected by neutral Higgs scalars in this class of models); (3) There are no doubly-charged Higgses in the theory; (4) There is no particular mechanism of mass generation for neutrinos, leaving the smallness of their masses unexplained.

The observables constraining the model will be the following: (a) EWPO (constraining VEVs and gauge couplings); (b) meson mixing and (semi-)leptonic decays (for $ V^{L} $, $ V^{R} $, Higgs masses, etc.); (c) and finally other observables as $ b \rightarrow s \gamma $ and the relation among the masses of up- and down-type quarks. The firs two categories will be discussed in the following sections.

\section{EWPO}

% Observables
Among the observables one can use to constrain models of New Physics (NP), EWPO are of particular
importance due to the accuracy reached by SM computations and experiments \cite{PDG}. The SM global fit of these observables shows good agreement between them, but some tension is present specially between the Forward-Backward asymmetry $ A_{FB} (b) $ and the Left-Right asymmetry $ A_{LR} (e) $. The observables we consider here are (a) Z-lineshape and asymmetries: $ A_{LR} (f) $, for $ f = e, \mu, \tau, c, b $, $ A_{FB} (f) $, for $ f = e, \mu, \tau, c, b $, the hadronic cross section at the Z-pole ($ \sigma^{0}_{\operatorname{had}} $), ratios of partial widths ($ R_{\ell} $, for $ \ell = e, \mu, \tau $, and $ R_{q} $, for $ q = c, b $) and the total width ($ \Gamma_{Z} $); (b) Mass ($ M_{W} $) and total width ($ \Gamma_{W} $) of the W; (c) Atomic parity violation of cesium and thallium.

In the SM, one usually parameterizes EWPO in terms of $ \mathcal{S} \equiv \{ m_{h}, m_{t}, \alpha_{s} (M_{Z}), \Delta \alpha, M_{Z} \} $:

\begin{equation}
\mathcal{X} = c_{0} + c_{1} \cdot L_{H} + c_{2} \cdot \Delta_{t} + c_{3} \cdot \Delta_{\alpha_{s}} + c_{4} \cdot \Delta^{2}_{\alpha_{s}} + c_{5} \cdot \Delta_{\alpha_{s}} \Delta_{t} + c_{6} \cdot \Delta_{\alpha} + c_{7} \cdot \Delta_{Z} ,
\end{equation}
\noindent
where $ \Delta_{t} = \left( \frac{m_{t}}{173.2 \, \operatorname{GeV}} \right)^{2} - 1 $, $ \Delta_{\alpha_{s}} = \frac{\alpha_{s} (M_{Z})}{0.1184} - 1 $, $ \Delta_{\alpha} = \frac{\Delta \alpha}{0.059} - 1 $, $ \Delta_{Z} = \frac{M_{Z}}{91.1876 \, \operatorname{GeV}} - 1 $ and \linebreak $ {L_{H} = \log \frac{m_{h}}{125.7 \, \operatorname{GeV}}} $. The values of the coefficients $ c_{i} $ for some observables ($ \Gamma_{Z} $, $ \sigma^{0}_{\operatorname{had}} $, $ R_{b, c} $), including 2-loop fermionic EW corrections, are given by \cite{Freitas}. Using the numerical program \textsf{Zfitter} \cite{Zfitter}, one can determine the values of the coefficients $ c_{i} $ of the other observables with a level of accuracy somewhat lower.
The fundamental parameters $ \mathcal{R} \equiv \{ \epsilon^2 \equiv \frac{\kappa^{2}}{\kappa^{2}_{R}}, c^{2}_R \equiv 1 - \frac{s^{2}_{W}}{1 - s^{2}_{W}} \left( \frac{g_{L}}{g_{R}} \right)^{2}, r \equiv \frac{\kappa_{2}}{\kappa_{1}}, w \equiv \frac{\kappa_{L}}{\kappa_{1}} \} $ are also present in the LRM, where $ s_{W} $ is the sine of the weak angle. The EWPO are given by $ \mathcal{X}_{LR} = \mathcal{X} + \delta \mathcal{X} $, where $ \delta \mathcal{X} $ is the Leading-Order (LO) correction to the SM and $ \frac{\delta \mathcal{X}}{\mathcal{X}} = \mathcal{O} (\epsilon^{2}) $. A similar treatment of EWPO can be found in \cite{Yuan}.

%\begin{minipage}[t]{0.8\textwidth}
\begin{figure}
\begin{minipage}[b]{0.4\linewidth}
\centering
\includegraphics[scale=0.3]{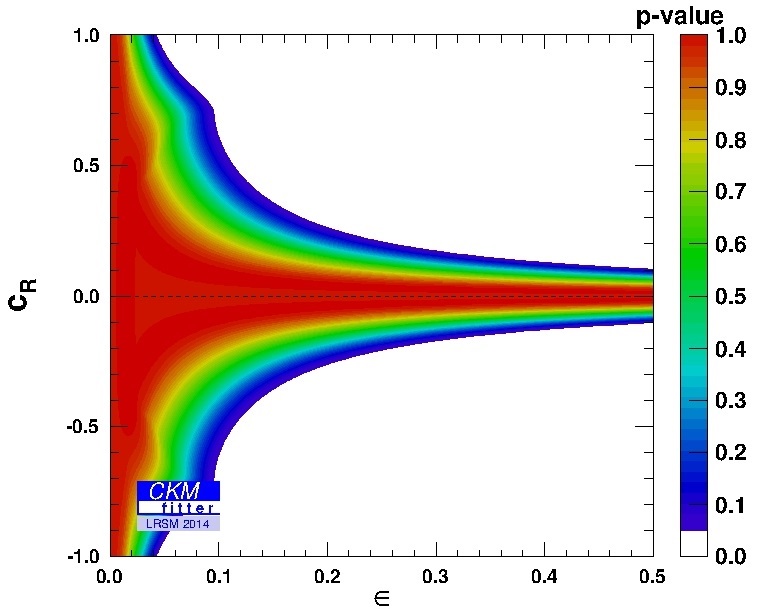}
\captionof{figure}{Correlation among $ \epsilon $ and $ c_R $. Perturbativity bounds, i.e. $ g^{2}_{R, B-L} < 4 \pi $, implying $ 0.1 \lesssim | c_{R} | < 1 $, are not shown.}
\label{fig:fig1}
\end{minipage}
\hspace{0.5cm}
\begin{minipage}[b]{0.55\linewidth}
		\begin{tabular}{|c|c|ccc|c|c|}
		\hline
		$ w $ & $ \epsilon^{2} $ & $ c_{R} $ & $ g_{R} $ & $ g_{B-L} $ & $ M_{Z'} $[TeV] & $ \chi^{2}_{min} $ \\
		\hline
		0 & 0.88 & 0.11 & 0.36 & 3.57 & 13.1 & 26.12 \\
		1 & 1.04 & 0.40 & 0.39 & 0.90 & 3.8 & 25.14 \\
		2 & 1.43 & 0.63 & 0.46 & 0.56 & 2.4 & 24.06 \\
		\hline
		\end{tabular}
		\captionof{table}{Central values of the parameters of LRM, according to EWPO fits performed at different values of $ w $. $ M_{W'} = 1.5 $~TeV is fixed. $ \epsilon^{2} $ is given in units $ 10^{-3} $.}
		\label{tab:changingw}
\end{minipage}
\end{figure}
%\end{minipage}

In order to combine the different EWPO and constrain the parameters $ \mathcal{S} \cup \mathcal{R} $, we use the \textsf{CKMfitter} \cite{CKMfitter} frequentist framework (with \textit{Range fit} treatment of systematic uncertainties). The correlated constraint among the scale of LRM, $ \epsilon $, and the size of $ c_R $ is seen in Figure \ref{fig:fig1}. We do not use bounds on masses coming from direct searches for
the $ W' $ boson, as the latter are tied to specific assumptions on the structure of the LRM couplings \cite{directsearches} and the analysis should be adapted to the more general framework considered here.

The constraints are not powerful enough to constrain $ c_{R} $, $ r $ and $ w $ independently at $ 1 \sigma $. The global fit of LRM is similar to the SM one: $ \chi^{2}_{\min} \vert_{SM} = 22.24 $ and $ \chi^{2}_{\min} \vert_{LRM} = 22.19 $. The agreement is improved for some observables (e.g. $ \sigma^{0}_{\operatorname{had}} $) at the expense of others (e.g. $ \Gamma_{Z} $), \cite{Luiz}. Though not constrained at $ 1 \sigma $, $ w $ has an impact on the fit, as seen in Table \ref{tab:changingw}. The fit prefers $ w > 0 $, though $ \chi^{2}_{\min} $ does not change by large amounts. Moreover, when $ w = 0 $, $ g_{B-L} $ reaches its perturbativity limit, $ g^{2}_{B-L} = 4 \pi $. The fact that $ w $ is pushed towards non-vanishing values is an interesting feature of EWPO, but it remains to be seen if the other sectors of the theory agree with this tendency.

\section{Neutral-meson mixing}

In order to further test the LRM, in particular the scale of the masses of the Higgses and the general structure of the $ V^R $ mixing matrix, we consider meson oscillation observables. The SM calculation of these observables consists of $ {W}^{\pm} {W}^{\pm} $, $ {W}^{\pm} {G}^{\pm} $ and $ {G}^{\pm} {G}^{\pm} $ box diagrams ($ {G}^{\pm} $ is the Goldstone associated to $ {W}^{\pm} $), and it is corrected at order $ \epsilon^2 $ by (a) new boxes $ {W}^{\pm} {W'}^{\pm} $ and $ {G}^{\pm} {W'}^{\pm} $; (b) $ W $ gauge boson/charged scalar boxes, $ {W}^{\pm} {H}^{\pm}_{1,2} $ and $ {G}^{\pm} {H}^{\pm}_{1,2} $; (c) FCNC introduced by $ H^{0}_{1,2} $ and $ A^{0}_{1,2} $ at tree-level; and (d) self-energy and vertex corrections to the FCNC, necessary for gauge invariance of the $ {W}^{\pm} {W'}^{\pm} $ box \cite{Pal}. Usually, the tree-level Higgs exchanges dominate over the other new contributions. In the triplet case (which is similar to the limit $ \kappa_{L} = 0 $), only the pair $ H^{0}_{1} $, $ A^{0}_{1} $ contributes. In the doublet case, the presence of contributions from other Higgses, $ H^{0}_{2} $ and $ A^{0}_{2} $, and different FCNC couplings when $ \kappa_{L} \neq 0 $, as suggested by the analysis of EWPO, means that the constraint from neutral meson mixing is less stringent, in particular on the mass of FCNC Higgses.

The general structure of the neutral meson mixing observables is $ \Delta m = \sum_{i} C^{q_{1} q_{2}}_{i} \eta^{q_{1} q_{2}}_{i} \langle O_{i} \rangle $, where $ i $ runs over the number of operators, and $ q_{1}, q_{2} $ are the flavors of the up-type quarks in the box, also related to the FCCC mixing matrices arising in $ C_{i} $. The Wilson coefficients $ C_{i} $ can be computed perturbatively by matching the low energy EFT (effective Hamiltonian) and the underlying theory (SM here), whereas the matrix element $ \langle O_{i} \rangle $ can be determined from lattice QCD; $ \eta $, collecting the short-distance QCD corrections are precisely known (up to NNLO) in the SM through the use of EFT. There is also a simplified method for computing the $ \eta $'s, described by Vysotskii \cite{Vysotskii}. Consider the LO diagrams (for instance, the $ W W $ box) with the addition of a gluon (corresponding to a two-loop integral on a gluon momentum and a quark momentum). Vysotskii's simplified method aims at extracting the main contributions to the short-distance coefficients by determining, at a first stage, the range of momenta of the gluon contributing to the leading-order QCD corrections. One then improves the result with the help of Renormalization Group Equations (RGE), resumming the gluon corrections thanks to the running of four-fermion operators. Finally, the quark momentum of the 2-loop integral is determined from the range of energies dominating the Inami-Lim functions (LO computation). The values of the short-distance QCD corrections in the SM for the $ K $ system using this method are given in Table \ref{tab:SMetas}, and they reproduce the values calculated from a systematic use of EFT \cite{Burasetas}.

For the LR operators, only calculations of the $ \eta $ for top-top box below $ \mathcal{O} (m_{t}) $ are known \cite{Burasgammas}. To derive constraints from meson mixing, we compute the remaining $ \eta $'s applying the procedure described by Vysotskii, extending what was done by \cite{Ecker}. We give our preliminary results in Table \ref{tab:LRetas}. The same approach was also employed by \cite{Bertolini}. More details of our calculations at LO, a possible NLO extension and the corresponding constraints on the parameters of the Doublet LRM from meson mixing will be given in ref. \cite{Luiz}.

\begin{table}
\centering
\begin{tabular}{c|ccc}
\hline
$ K \overline{K} $, LO & $ \eta_{tt} $ & $ \eta_{cc} $ & $ \eta_{ct} $ \\
\hline
Vysotskii \cite{Vysotskii} & 0.60 & 0.92 & 0.34 \\
\hline
systematic EFT \cite{Burasetas} & 0.612 & 1.12 & 0.35 \\
\hline
\end{tabular}
\caption{SM short-distance QCD corrections at LO showing a comparison between Vysotskii's prescription and a systematic use of Effective Field Theory (EFT). Flavor thresholds are taken into account.}\label{tab:SMetas}
\end{table}

\begin{table}
\begin{tabular}{|c|c|c|c|c|}
\hline
LO & $ \overline{\eta}^{K \overline{K}}_{tt} $ & $ \overline{\eta}^{K \overline{K}}_{cc} $ & $ \overline{\eta}^{K \overline{K}}_{ct} $ & $ \overline{\eta}^{B \overline{B}}_{tt} $ \\
\hline
$ W^{\pm} {W'}^{\pm} $, $ W^{\pm} H^{\pm} $ & 2.89 & 0.78 & 1.50 & 2.19 \\
\hline
$ G^{\pm} {W'}^{\pm} $ & 2.89 & 0.92 & 1.50 & 2.19 \\
\hline
\end{tabular}
\begin{tabular}{|c|c|c|c|c|}
\hline
LO & $ \overline{\eta}^{K \overline{K}}_{tt} $ & $ \overline{\eta}^{K \overline{K}}_{cc} $ & $ \overline{\eta}^{K \overline{K}}_{ct} $ & $ \overline{\eta}^{B \overline{B}}_{tt} $ \\
\hline
$ G^{\pm} H^{\pm} $ & 2.89 & 0.31 & 0.41 & 2.18 \\
\hline
tree-level FCNC & 2.15 & 0.58 & 1.12 & 1.63 \\
\hline
\end{tabular}
\caption{Preliminary results for the short-distance QCD corrections at LO to the LRM, using the Vysotskii's procedure described briefly in the text. Flavor thresholds are taken into account. The $ \overline{\eta} $'s are dependent on the hadronisation scale: $ \mu_{\operatorname{had}} = 2 $~GeV is taken for the $ K $ system and $ \mu_{\operatorname{had}} = 4 $~GeV for the $ B $ systems.}\label{tab:LRetas}
\end{table}

\pagebreak
\section{Outlook}

% Model constrained, EWPO, meson oscillations, pariy
We reanalyze a version of LRM where the spontaneous breakdown of the LR gauge group is triggered by doublet rather than triplet representations. This changes the structure of the model and introduces a new degree of freedom, the VEV $ \kappa_{L} $ of a $ SU(2)_{L} $ doublet field. Our first concern was to set constraints in this model using EWPO. Our analysis shows that these observables impose a correlation between the scale of the left-right symmetry breaking, which occurs at the scale of several TeV scale, and the size of the couplings.

We are presently analyzing meson mixing observables in order to constrain the remaining parameters of the model. These observables are of great impact because LRM introduces FCNC through new heavy scalars. The structure of the corrections from LRM is different, in the doublet and triplet cases, and they require a good knowledge of short-distance QCD corrections, which can be large, as seen in Table \ref{tab:LRetas}. We aim at investigating the consequences of this new scalar sector, especially for FCNC, which are sensitive to $ \kappa_{L} $. A joint fit of EWPO and meson mixing observables, together with leptonic and semileptonic decays, is in progress \cite{Luiz}.


\begin{thebibliography}{ieeetr}

\bibitem{History}
	J.C. Pati and A. Salam, Phys. Rev. D10 (1974) 275-289 [Erratum-ibid. D11, 703 (1975)].
	R.N. Mohapatra and J.C. Pati, Phys. Rev. D11 (1975) 2558.
	R.N. Mohapatra and J.C. Pati, Phys. Rev. D11 (1975) 566-571.
	G. Senjanovic and R.N. Mohapatra, Phys. Rev. D12 (1975) 1502.
	G. Senjanovic, Nucl. Phys. B153 (1979) 334

\bibitem{Deshpande}
	N.G. Deshpande, J.F. Gunion, B. Kayser, F. Olness, Phys. Rev. D 44, No. 3, 1991

\bibitem{Buras}
	M. Blanke, A.J. Buras, K. Gemmler and T. Heidsieck, JHEP 1203 (2012) 024
	%hep-ph/1111.5014v2 (2012)

\bibitem{Mohapatra}
	Y. Zhang, H. An, X. Ji, R.N. Mohapatra, Nucl. Phys. B 802 (2008) 247-279
	%hep-ph/0712.4218v1 (2007)

\bibitem{Basecq}
	P. Ball et al., Nucl. Phys. B 572 (2000) 3.
	P. Langacker et al., Phys. Rev. D 40, No. 5, 1569 (1989)
	%P. Ball, J.-M. Frere and J. Matias, Nucl. Phys. B 572 (2000) 3-35.
	%P. Langacker and S.U. Sankar, Phys. Rev. D 40, No. 5, 1569 (1989)

%\bibitem{Barenboim}
%	G. Barenboim, M. Gorbahn, U. Nierste, M. Raidal, hep-ph/0107121v3	

\bibitem{Maiezza}
	A. Maiezza, M. Nemevsek, F. Nesti, G. Senjanovic, Phys. Rev. D 82 (2010) 055022
	%hep-ph/1005.5160v1

\bibitem{PDG}
	K.A. Olive et al. (Particle Data Group), Chin. Phys. C, 38, 090001 (2014) 

\bibitem{Freitas}
	A. Freitas, JHEP 1404 (2014) 070
	%hep-ph/1401.2447v1

\bibitem{Zfitter}
	Zfitter Group (A.B. Arbuzov), Comput. Phys. Commun. 174 (2006) 728-758
	%hep-ph/0507146v1

\bibitem{Yuan}
	K. Hsieh, K. Schmitz, J.-H. Yu, C.-P. Yuan, Phys. Rev. D 82, 035011 (2010)

\bibitem{CKMfitter}
	CKMfitter Group (J. Charles et al.), Eur. Phys. J. C41, 1-131 (2005)

\bibitem{directsearches}
	ATLAS Collaboration (G. Aad et al.), arXiv:1108.6311.
	CMS Collaboration (S. Chatrchyan et al.), http://cdsweb.cern.ch/record/1370086?ln=en.
	G. Altarelli et al., Z. Phys. C45, 109-121 (1989)
	% G. Altarelli, B. Mele, M. Ruiz-Altaba

\bibitem{Burasetas}
	Buchalla, Buras and Lautenbacher, Rev. Mod. Phys., Vol. 68, No. 4, October 1996

\bibitem{Pal}
	J. Basecq, L.-F. Li, P.B. Pal, Phys. Rev. D 32, No. 1, 1985

%\bibitem{NNLO}
%	J. Brod, M. Gorbahn, Phys. Rev. Lett. 108, 121801 (2012).
%	J. Brod, M. Gorbahn, Phys. Rev. D 82, 094026 (2010)

\bibitem{Burasgammas}
	A.J. Buras, S. Jager and J. Urban, Nucl. Phys. B 605 (2001) 600-624
	%hep-ph/0102316

\bibitem{Vysotskii}
	M.I. Vysotskii, Yad. Fiz. 31 (1980) 1535 [Sov. J. Nucl. Phys. 31 (1980) 797]

%\bibitem{Herrlichetacc}
%	S. Herrlich, U. Nierste, Nucl. Phys. B419 (1994) 292-322

%\bibitem{Herrlichetact}
%	S. Herrlich, U. Nierste, hep-ph/9604330

\bibitem{Ecker}
	G. Ecker and W. Grimus, Nucl. Phys. B 258, 328 (1985)

%\bibitem{MohapatraPotential}
%	R.N. Mohapatra, F.E. Paige, D.P. Sidhu, Phys. Rev.D, Vol. 17, nb. 9, 1 may 1978

\bibitem{Bertolini}
	S. Bertolini, A. Maiezza and F. Nesti, Phys. Rev. D 89 (2014) 095028
	%hep-ph/1403.7112v1

\bibitem{Luiz}
	V. Bernard, S. Descotes-Genon and L. Vale Silva, work in preparation

\end{thebibliography}
\end{document}